\documentclass[conference]{IEEEtran}

\IEEEoverridecommandlockouts
\usepackage[T1]{fontenc}
\usepackage{graphicx}
\usepackage{pifont}
\usepackage{threeparttable}
\usepackage{cite}
\usepackage{url}
\usepackage[utf8]{inputenc}
\usepackage{cite}  
\usepackage{hyperref} 
\usepackage{comment}
\usepackage{amsmath} 
\usepackage{bm}
\usepackage{amssymb} 
\usepackage{booktabs} 

\def\BibTeX{{\rm B\kern-.05em{\sc i\kern-.025em b}\kern-.08em
    T\kern-.1667em\lower.7ex\hbox{E}\kern-.125emX}}
\begin{document}

\title{LABNet: A Lightweight Attentive Beamforming Network for Ad-hoc Multichannel Microphone Invariant Real-Time Speech Enhancement}

\author{Haoyin Yan,Jie Zhang, Chengqian Jiang,Shuang Zhang\\
NERC-SLIP, University of Science and Technology of China (USTC), Hefei, China\\

}
\maketitle
\begin{abstract}
Multichannel speech enhancement (SE) aims to restore clean speech from noisy measurements by leveraging spatiotemporal signal features. In ad-hoc array conditions, microphone invariance (MI) requires systems to handle different microphone numbers and array geometries. From a practical perspective, multichannel recordings inevitably increase the computational burden for edge-device applications, highlighting the necessity of lightweight and efficient deployments. In this work, we propose a lightweight attentive beamforming network (LABNet) to integrate MI in a low-complexity real-time SE system. We design a three-stage framework for efficient intra-channel modeling and inter-channel interaction. A cross-channel attention module is developed to aggregate features from each channel selectively. Experimental results demonstrate our LABNet achieves impressive performance with ultra-light resource overhead while maintaining the MI, indicating great potential for ad-hoc array processing.The code is available:https://github.com/Jokejiangv/LABNet.git
\end{abstract}

\section{Introduction}

In real-world scenarios, environmental noises inevitably deteriorate the instrumental quality and intelligibility of speech signals. Multichannel recordings contain exploitable spatial information to distinguish between target speech and noises, implying greater speech enhancement (SE) potential than single-channel recordings~\cite{7805139}. With the development of deep learning techniques, deep neural network (DNN) based multichannel SE methods (also called neural beamformers) have emerged as front-end modules in many applications, e.g.,  human-machine interaction and audio-video conferencing~\cite{7886357,9064910,9730152}.

Numerous existing approaches have been developed for multichannel SE. For instance, EaBNet~\cite{9746432} proposes a causal neural beamformer paradigm called embedding and beamforming, which utilizes two modules to learn 3-D embedding tensors and derive the beamforming weights, respectively. In~\cite{10095509}, a multi-cue fusion network named McNet sequentially integrates four modules to exploit full-band spatial, narrowband spatial, sub-band spectral and full-band spectral features. 
These models are usually customized for fixed arrays, which are characterized by pre-defined microphone numbers and/or array geometries. In contrast, ad-hoc microphone arrays involve arbitrary array topologies, offering considerable flexibility in dynamic real-world scenarios, e.g., wearable device interactions and smart home applications. Providing invariance to array configurations is necessary for ad-hoc array processing. Fixed array methods can achieve geometry invariance by training with randomly placed microphones, but the obtained models must be modified and retrained to handle various numbers of microphones. It is promising to design a model for arbitrary array configurations with a single training session, so as to quickly adapt to new cases, i.e., {\it microphone invariance}.

Conventional statistics-based beamformers are typically formulated as optimization problems irrelevant to the microphone number and array geometry, conforming to microphone invariance. Some studies~\cite{7471664,8462430,8070987} apply DNNs to perform pre-enhancement on each channel independently and then perform conventional beamforming, such as minimum variance distortionless response beamforming (MVDR) or multichannel Wiener filtering (MWF).
Nevertheless, unreliable first-stage results will interfere the output due to the separation of these two stages. When DNN-based models are considered to handle the multi-input single-output (MISO) problem in an end-to-end manner, some special paradigms must be adopted to retain microphone invariance. With a two-stage design, the filter-and-sum network (FaSNet)~\cite{9003849} first performs enhancement on a selected reference channel, and the beamforming filters of the remaining channels are then estimated based on pair-wise cross-channel features between the reference channel and others. All filtered signals are summed to produce the final output. In~\cite{9054177}, the transform-average-concatenate (TAC) method is proposed to aggregate the information from all microphones, improving the performance of FaSNet. Before performing filter-and-sum, the delay-filter-and-sum network (DFSNet)~\cite{kovalyov23_interspeech} delays time-domain waveforms toward the speech source direction by using integer and fractional delay finite impulse response (FIR) filters. Although these methods can achieve microphone invariance, it is meaningful to develop a more lightweight and effective model for the deployment on low-resource devices. 
\begin{figure*}
    \centering
    \includegraphics[width=1.0\textwidth]{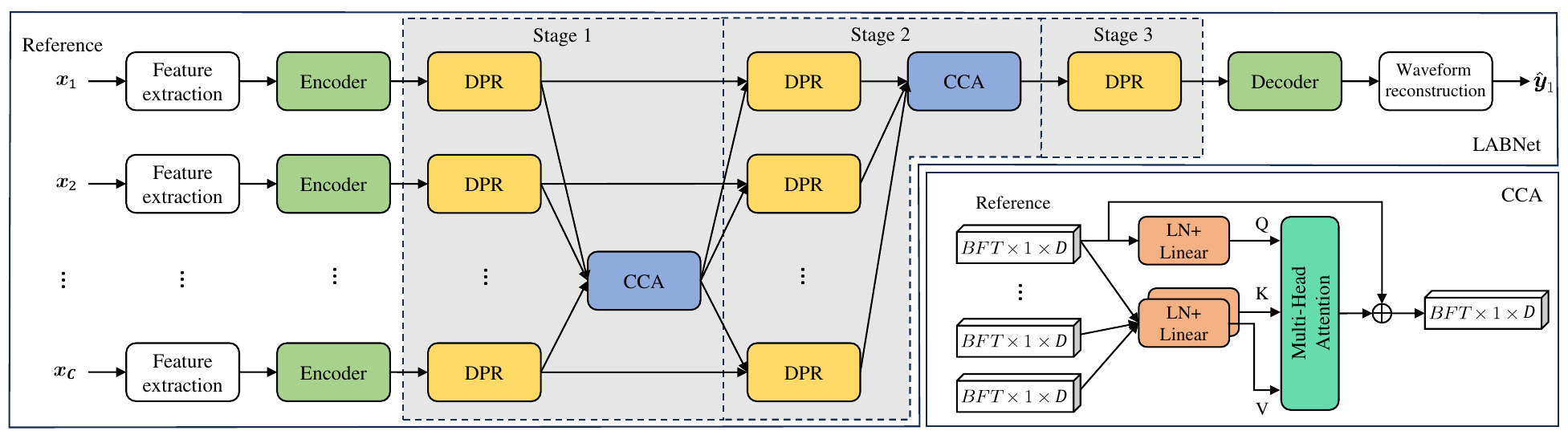}
    \caption{The overall architecture of our proposed LABNet and the details of the CCA module. The main framework consists of feature extraction, encoder, three-stage paradigm, decoder and waveform reconstruction.}
    \label{fig:architecture}
\end{figure*}

In this work, we therefore propose a \textbf{L}ightweight \textbf{A}ttentive \textbf{B}eamforming \textbf{Net}work (LABNet) for multichannel real-time speech enhancement in ad-hoc scenarios. To efficiently aggregate intra-channel and inter-channel information while retaining the microphone invariance, we design a three-stage framework within the network. In the channel-wise processing stage, hidden representations from each channel are processed independently and then fused in the reference channel. In the pair-wise alignment stage, we extract pair-wise features by concatenating the reference channel with others, which helps eliminate delays caused by different source-microphone distances.
After fusing information from all channels, the reference channel representation is refined in the post refinement stage. Additionally, we develop a cross-channel attention module to aggregate information from all channels selectively, which is suitable for any number of microphones. Experimental results show that the proposed LABNet surpasses other baselines in SE performance with ultra-light resource overhead. Ablation studies validate the effectiveness of the involved modules.

\section{Methodology}

\subsection{Problem Formulation}

Given time-domain speech source $\bm{y}$ and $C$ microphones with an arbitrary geometry, the noisy signal received by the $c$-th microphone is defined as 
\begin{align}
    \bm{x}_c = \bm{y}_c + \bm{n}_c = \bm{y} \circledast \bm{h}_c + \bm{n}_c, \quad c=1,2,...,C,
\end{align}
where $\bm{y}_c$ denotes the received reverberant speech signal, $\circledast$ the convolution operator, $\bm{h}_c$ the room impulse response (RIR) from the source position to microphone $c$, and $\bm{n}_c$ the additive noise component at the $c$-th microphone, respectively. We consider the scenario with one speech source and multiple noise sources, indicating $\bm{n}_c$ is the combination of multiple reverberant noise signals. Speech signals received by each microphone inevitably encounter different delays due to their arbitrary spatial locations. To avoid ambiguity of target signals during training, selecting a reference channel is necessary. We regard the first channel as the reference without loss of generality, which can be chosen by a more sophisticated method like~\cite{Zhang2021Study}, the goal of this work is thus to estimate the reverberant speech signal $\bm{y}_1$ from noisy recordings $\{\bm{x}_c\}^C_{c=1}$.

\subsection{Overall Architecture}

The network architecture of our LABNet is extended from the single-channel real-time SE model LiSenNet~\cite{yan2024lisennetlightweightsubbanddualpath}, as shown in Fig. \ref{fig:architecture}. For each channel, we extract the power-compressed magnitude spectrum and phase differences (PDs) of the noisy signal in the time-frequency domain using the short-time Fourier transform (STFT), which are concatenated as a 3-dimensional network input. The encoders, composed of several convolution blocks with shared parameters, map inputs from all channels to high-dimensional embeddings independently. The dual-path recurrent (DPR) module utilizes the dual-path architecture~\cite{9054266} to capture time and frequency dependencies efficiently, which employs a bidirectional GRU (Bi-GRU) for frequency modeling, an unidirectional GRU for temporal modeling, and a ConvGLU~\cite{10655567} for hidden-dimension information fusion. We integrate the DPR module with the proposed cross-channel attention (CCA) module within a three-stage processing framework, facilitating comprehensive intra-channel feature extraction and inter-channel modeling, see Section \ref{sec:cca}. The decoder shares a similar architecture with the encoder, decoding the single-channel hidden representation into the magnitude spectrum mask of the reference channel. The Griffin-Lim Algorithm (GLA)~\cite{1164317,6701851} is leveraged for phase refinement based on the estimated magnitude spectrum. Combining the estimated magnitude and phase spectrum, we restore the estimated clean waveform $\hat{\bm{y}}_1$ by the inverse STFT (iSTFT). Multiple loss functions are considered to optimize the network, including the metric loss and the mean squared error (MSE) on the power-compressed magnitude spectrum and complex spectrum. The metric loss is calculated using a discriminator~\cite{pmlr-v97-fu19b} trained to predict the perceptual evaluation of speech quality (PESQ)~\cite{941023} score of the estimated speech. Please refer to~\cite{yan2024lisennetlightweightsubbanddualpath} for more details on the feature extraction, waveform reconstruction, loss functions and network modules.

\subsection{Cross-Channel Attention (CCA)}
\begin{table*}[]
    \centering
    \caption{Performance comparison on the test set with 6 microphones. \textbf{T-F} and \textbf{T} denote the time-frequency domain and time domain methods, respectively. \textbf{MI} indicates whether each approach conforms to microphone invariance. 
    }
    \begin{tabular*}{\textwidth}{@{\extracolsep{\fill}}l|cccccc|ccccc}
    \toprule
    Method & Domain & Causal & MI & Para. & MACs & Latency & PESQ & STOI & CSIG & CBAK & COVL \\
    \midrule
    Noisy & - & - & - & - & - & - & 1.48 & 0.81 & 3.10 & 2.09 & 2.27 \\
    Oracle MVDR & T-F & \ding{56} & \ding{52} & - & - & - & 2.18 & 0.78 & 3.87 & 2.33 & 3.06 \\
    EaBNet~\cite{9746432} & T-F & \ding{52} & \ding{56} & 2.8M & 7.4G & 20ms & 2.76 & 0.93 & 4.40 & \textbf{3.33} & 3.64 \\
    McNet~\cite{10095509} & T-F & \ding{52} & \ding{56} & 1.9M & 30.1G & 32ms & 2.82 & \textbf{0.94} & 4.39 & \textbf{3.33} & 3.67 \\
    FaSNet~\cite{9003849} & T & \ding{52} & \ding{52} & 2.8M & 9.1G & 220ms & 2.18 & 0.90 & 3.66 & 2.59 & 2.93 \\
    FaSNet-TAC~\cite{9054177} & T & \ding{52} & \ding{52} & 2.3M & 11.7G & 120ms & 2.25 & 0.91 & 3.70 & 2.67 & 2.99 \\
    DFSNet\tnote{*}~\cite{kovalyov23_interspeech} & T & \ding{52} & \ding{52} & 549k & 1.7G & \textbf{4.5ms} & 2.01 & 0.87 & 3.57 & 2.45 & 2.79 \\
    \midrule
    LABNet & T-F & \ding{52} & \ding{52} & \textbf{52k} & \textbf{0.316G} & 64ms & \textbf{2.92} & 0.93 & \textbf{4.42} & 3.24 & \textbf{3.70} \\
    \bottomrule
    \end{tabular*}


    \vspace{-0.5em}
    \label{tab:comparision}
\end{table*}
\label{sec:cca}

Instead of employing the simple average to aggregate features from all channels~\cite{9054177}, we adopt the attention mechanism to facilitate selective integration of multichannel information, as shown at the lower right corner of Fig. \ref{fig:architecture}. 
The CCA module processes the encoded hidden representations $\bm{h}^{in} \in \mathbb{R}^{BFT \times C \times D}$ from all channels, generating a single-channel representation $\bm{h}^{out} \in \mathbb{R}^{BFT \times 1 \times D}$, where $B$, $F$, $T$ and $D$ denote the batch size, number of frequency bins, frame amount and hidden dimensions, respectively. Let $\bm{h}_1^{in} \in \mathbb{R}^{BFT \times 1 \times D}$ indicate the reference hidden representation. The query ($Q$) with unit sequence length is derived by applying layer normalization (LN) followed by a linear transformation (Linear) to $\bm{h}_1^{in}$, while the key ($K$) and value ($V$) with sequence length $C$ are similarly computed from the complete input $\bm{h}^{in}$ through similar processing steps. The output can be formulated as
\begin{align}
    \bm{h}^{out} = {\rm MHA}(Q, K, V) + \bm{h}_1^{in},
\end{align}
where ${\rm MHA}(\cdot)$ denotes the multi-head attention~\cite{NIPS2017_3f5ee243}. The advantages of the CCA module are twofold: the reference channel can adaptively aggregate valuable information from other channels; it reduces the number of channels to one, conforming to the MISO paradigm. In addition, since the attention mechanism is independent of sequence length and permutation, the microphone invariance can be guaranteed.

\subsection{Three-Stage Framework}

Based on the DPR and CCA modules, we design a three-stage framework for efficient intra-channel and inter-channel modeling. 
During the initial channel-wise processing stage (Stage 1), given the encoded representations of all channels $\bm{h}^e = \left[ \bm{h}_1^e, \bm{h}_2^e, \cdot\cdot\cdot, \bm{h}_C^e \right] \in \mathbb{R}^{B \times C \times D \times T \times F}$, we apply DPR modules with shared parameters to each channel before utilizing a CCA module to produce $\bm{h}_r^{s1}$, which can be formulated as
\begin{align}
    \bm{h}_c^{s1} &= {\rm DPR}(\bm{h}_c^{e}), \quad c=1,2,...,C, \\
    \bm{h}_r^{s1} &= {\rm CCA}(\left[ \bm{h}_1^{s1}, \bm{h}_2^{s1}, \cdot\cdot\cdot, \bm{h}_C^{s1} \right]),
\end{align}
where $\bm{h}_c^{s1}$ indicates the hidden representation of the $c$-th channel after first-stage processing. Appropriate reshaping operations are applied to connect different modules if necessary. The DPR module captures time and frequency dependencies within each channel, while the CCA module preserves temporal information of the reference channel and integrates spatial cues from other channels. Since arbitrary microphone locations can lead to various temporal delays of the source signal, we utilize $\bm{h}_r^{s1}$ to perform alignment between the reference channel and others, resulting in the pair-wise alignment stage (Stage 2). Specifically, $\bm{h}_r^{s1}$ is concatenated with the hidden representation of each channel along the hidden dimension, followed by being processed by a parameter-shared linear layer and DPR module to reduce dimensionality and achieve adaptive temporal alignment, respectively. We further employ a CCA module to transform multichannel input into single-channel output $\bm{h}_1^{s2}$ as
\begin{align}
    \bm{h}_c' &= {\rm DPR}({\rm Linear}(\left[ \bm{h}_r^{s1}: \bm{h}_c^{s1} \right])), \quad c=1,2,...,C, \\
    \bm{h}_1^{s2} &= {\rm CCA}(\left[ \bm{h}_1', \bm{h}_2', \cdot\cdot\cdot, \bm{h}_C' \right]),
\end{align}
where $\bm{h}_c'$ denotes the intermediate representation and $\left[ 
\cdot:\cdot \right]$ indicates the concatenation operation. Finally, in the post refinement stage (Stage 3), we utilize a DPR module with powerful time-frequency modeling capabilities to further refine $\bm{h}_1^{s2}$ as
\begin{align}
\bm{h}_1^{s3} = {\rm DPR}(\bm{h}_1^{s2}).
\end{align}
After three-stage processing, the decoder produces the magnitude spectrum mask based on $\bm{h}_1^{s3}$. The impact of each stage will be analyzed in experiments.

\section{Experiments}
\subsection{Dataset}
We generate a simulation dataset using clean utterances from WSJ0~\cite{wsj0} corpus and noise recordings from WHAM!~\cite{wichern19_interspeech} dataset, resulting in 12776 clips (24.9 hours) for training, 1206 clips (2.2 hours) for evaluation and 651 clips (1.5 hours) for testing. Each simulated recording contains one target speech source accompanied by multiple noise sources, with the number of noise sources randomly varying between 1 and 3. The overall signal-to-noise ratio (SNR) is uniformly sampled between -5 and 15 dB. We employ a configuration of 6 microphones during training and validation, while there are 12 microphones in the test set to assess model performance in case of an unseen number of microphones. For each recording, we simulate an acoustic environment by randomly generating room dimensions with length and width uniformly distributed between 5 and 10 meters, and height varying from 3 to 4 meters. The sources and microphones are randomly located with the constraint of being at least 0.5 meters away from the walls. The reverberation time (T60) randomly ranges from 0.1 to 0.5 seconds, and the room impulse responses (RIRs) are generated by the image method~\cite{Allen1976ImageMF} implemented by the gpuRIR toolbox~\cite{Diaz_Guerra_2020}. All recordings are sampled at 16 kHz.

\subsection{Implementation Details}

During training, we dynamically shuffle channel permutations and randomly select the number of channels between 1 and 6 to enhance data diversity. The utterances in the training set are dynamically sliced into 4-second segments to form batches. We perform STFT using a Hann window with a length of 512 (32ms) and a shift of 256 (16ms). The hyper-parameter setup of the encoder, DPR, and decoder module in our model follows the configurations in~\cite{yan2024lisennetlightweightsubbanddualpath}. The number of attention heads in the CCA module is set to 4. We employ one iteration for GLA, resulting in a total algorithm latency of 64ms. The model is trained using the AdamW~\cite{loshchilov2019decoupled} optimizer for 100 epochs, with the gradient clipping factor set to 5.0. The learning rate starts from 5e-4 and decays at a factor of 0.98 every epoch.


\subsection{Experimental Results}

Our LABNet is compared with several multichannel SE methods, including a conventional approach (Oracle MVDR\footnote{https://pypi.org/project/beamformers}), fixed array approaches (EaBNet~\cite{9746432}, McNet\cite{10095509}), and ad-hoc array approaches (FaSNet~\cite{9003849}, FaSNet-TAC~\cite{9054177}, DFSNet~\cite{kovalyov23_interspeech}). Additionally, since the LABNet can also process monaural recordings during inference, we compare it with several models designed for monaural SE, including CRN~\cite{tan18_interspeech}, DCCRN~\cite{hu20g_interspeech}, FullSubNet~\cite{9414177}, Fast FullSubNet~\cite{hao2023fastfullsubnet}, and LiSenNet~\cite{yan2024lisennetlightweightsubbanddualpath}. Five commonly-used objective evaluation metrics are chosen to evaluate the enhanced speech quality, including PESQ, short-time objective intelligibility (STOI)~\cite{5713237}, and three composite mean opinion score (MOS) based measures~\cite{4389058}, i.e., CSIG (the MOS of signal distortion), CBAK (the intrusiveness of background noise), and COVL (the overall effect). For all considered metrics, higher values indicate a better performance.

\begin{figure}
    \centering
    \includegraphics[width=0.9\columnwidth]{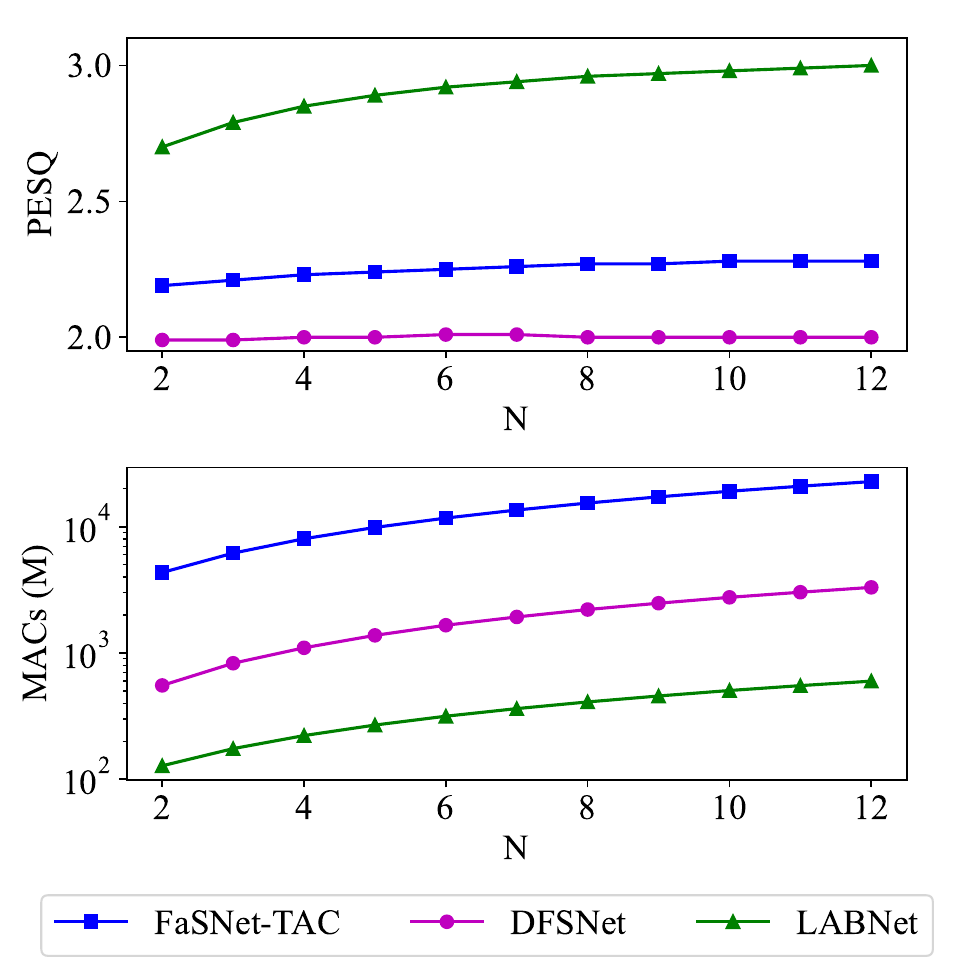}
    \caption{The PESQ and MACs of FaSNet-TAC, DFSNet and our LABNet in terms of different numbers of microphones.}
    \vspace{-0.5em}
    \label{fig:pesq_macs}
\end{figure}

First, in Table~\ref{tab:comparision} we report the SE performance obtained on the test set with 6 microphones. The Oracle MVDR utilizes the ground-truth spatial covariance matrices of the noisy signal, providing the upper-bound performance of MVDR-based beamforming techniques. EaBNet, McNet and our LABNet outperform Oracle MVDR in terms of all metrics, indicating the superiority of DNN-based methods in multichannel SE. In comparison with EaBNet and McNet, our model exhibits comparable performance with significantly fewer parameters (Para.) and multiply-accumulate operations (MACs). Furthermore, our proposed LABNet offers the crucial advantage of microphone invariance, showing superior adaptability and flexibility in practical applications. When compared with FaSNet, FaSNet-TAC and DFSNet, our ultra-light model achieves an obvious superiority in the SE performance. FaSNet and FaSNet-TAC adopt the dual-path architecture~\cite{9054177}, which performs segmentation on temporal frames, thus increasing the algorithm latency. We utilize the ground-truth time difference of arrival (TDOA) to align time-domain waveforms in DFSNet, which is difficult to obtain in practice. Calculating TDOA increases the complexity of the SE system and inaccurate estimates will degrade the performance. Fig.~\ref{fig:pesq_macs} visualizes the PESQ and MACs of the methods that can conform to microphone invariance in terms of different numbers of microphones. Although our model is trained with a maximum of 6 microphones, it can be generalized to accommodate more channels, which means higher computational complexity. Since increasing microphones brings decreasing additional benefits, the number of channels can be flexibly adjusted to achieve a trade-off between the resource usage and SE performance. Our LABNet maintains microphone invariance and achieves promising SE performance with an ultra-light resource overhead and acceptable algorithm latency.

In Table~\ref{tab:comparison2}, we compare the single-channel version of the proposed LABNet with several state-of-the-art (SOTA) monaural SE approaches, which are trained by randomly and dynamically selecting a channel from each 6-channel recording during training. For fairness, the number of DPR modules is set to 3 in LiSenNet that is consistent with our LABNet. Compared with CRN, DCCRN, FullSubNet, and Fast FullSubNet, the proposed model shows comparable or superior capabilities in SE, but significantly reduces the resource consumption, e.g., requiring only 0.9\% parameters and 0.3\% MACs compared to FullSubNet. Our LABNet also exhibits a similar performance as the monaural version LiSenNet with a marginally increased resource usage, which is caused by the CCA module. However, our model can achieve better SE performance by employing spatial information from multichannel recordings, albeit at the cost of higher computational complexity, demonstrating valuable extensibility in practice.

Finally, we conduct an ablation study using 6 microphones in the test set to verify the efficacy of our three-stage framework in Table~\ref{tab:ablation}. The removal of either channel-wise processing stage (w/o Stage 1), pair-wise alignment stage (w/o Stage 2), or post refinement stage (w/o Stage 3), leads to a certain performance degradation. The most significant effect is caused by that of Stage 1, highlighting the necessity of performing intra-channel modeling and integrating spatial cues from all channels, which can provide crucial supports for subsequent stages. Replacing our CCA module with the TAC (w/ TAC) also leads to a performance drop, showing the superiority of the CCA module in selective integration of inter-channel information.

\section{Conclusion}
\label{sec:conc}

In this work, we proposed a lightweight beamforming network called LABNet for multichannel real-time SE. We designed a three-stage framework that includes channel-wise processing, pair-wise alignment and post refinement to effectively model intra-channel patterns and capture inter-channel interactions. The CCA module was developed to adaptively aggregate features from each channel using attention mechanism. Our network paradigm guaranteed the capability to handle arbitrary microphone numbers and array geometries. Experiments on a simulated dataset demonstrated that our model can achieve better performance than ad-hoc array approaches and offered greater extensibility than fixed array methods and monaural SE methods. Remarkably, LABNet achieved these advancements with minimal resource overhead, indicating the great potential for ad-hoc array signal processing in practice. 
Future work should cover adaptively selecting the reference channel and reducing the algorithm latency for latency-sensitive applications.

\begin{table}[]
    \centering
    \caption{Comparison with SOTA monaural SE approaches on the test set using one microphone.}
    \resizebox{\columnwidth}{!}{
        \begin{tabular}{l|cc|ccc}
        \toprule
        Method & Para. & MACs & PESQ & STOI & COVL \\
        \midrule
        Noisy & - & - & 1.48 & 0.81 & 2.27 \\
        CRN~\cite{tan18_interspeech} & 17.6M & 2.6G & 2.00 & 0.88 & 2.97 \\
        DCCRN~\cite{hu20g_interspeech} & 3.7M & 14.4G & 2.33 & \textbf{0.90} & 3.16 \\
        FullSubNet~\cite{9414177} & 5.6M & 30.9G & 2.45 & \textbf{0.90} & \textbf{3.34} \\
        Fast FullSubNet~\cite{hao2023fastfullsubnet} & 6.8M & 4.1G & 2.17 & 0.89 & 3.09 \\
        LiSenNet~\cite{yan2024lisennetlightweightsubbanddualpath} & \textbf{48k} & \textbf{72M} & \textbf{2.51} & 0.88 & 3.28 \\
        \midrule
        LABNet & 52k & 80M & \textbf{2.51} & 0.88 & 3.29 \\
        \bottomrule
        \end{tabular}
    }
    \vspace{-0.5em}
    \label{tab:comparison2}
\end{table}

\begin{table}[]
    \centering
    \caption{Ablation study on the test set with 6 microphones.}
    \resizebox{\columnwidth}{!}{
        \begin{tabular}{l|cc|ccc}
        \toprule
        Method & Para. & MACs & PESQ & STOI & COVL \\
        \midrule
        LABNet & 52k & 316M & \textbf{2.92} & \textbf{0.93} & \textbf{3.70} \\
        \midrule
        w/o Stage 1 & 39k & 203M & 2.79 & 0.92 & 3.57 \\
        w/o Stage 2 & \textbf{38k} & \textbf{197M} & 2.83 & 0.92 & 3.60 \\
        w/o Stage 3 & 41k & 299M & 2.85 & 0.92 & 3.64 \\
        w/ TAC & 52k & 308M & 2.89 & 0.92 & 3.67 \\
        \bottomrule
        \end{tabular}
    }
    \vspace{-0.5em}
    \label{tab:ablation}
\end{table}
\vfill\pagebreak



\end{document}